# Lately Exposed Amorphous Water Ice on Comet 49P/Arend-Rigaux

B Sivaraman[a,*], V Venkataraman[b], A Kalyaan[c], S Arora[b,$], S Ganesh[b]

[a] Space and Atmospheric Sciences Division, [b] Astronomy and Astrophysics Division and [c] Planetary Exploration Division, Physical Research Laboratory, Ahmedabad.

Abstract

Comet 49P/ Arend-Rigaux, thought to be a low activity comet since the 1980's was found to be active in its recent apparitions. Recent analysis of the data obtained from Spitzer observation of the comet in 2006 compared with laboratory spectra has revealed amorphous water ice on the surface. In addition, in 2012 a jet was found to appear during its subsequent perihelion passage as witnessed during an observation carried out on 26th March 2012 using the PRL telescope at Mt. Abu. This confirms recent activity of Comet 49P/Arend-Rigaux due to the volatile subsurface materials exposed after several passages close to the Sun. Our result confirms the subsurface ices on cometary nuclei and insists for more observations for a better understanding.

[*]Corresponding author email: bhala@prl.res.in
[$]Presently at Astronomy and Astrophysics Division, Physical Research Laboratory, Ahmedabad, India.

## Introduction

Comet Arend- Rigaux, suspected of being a near extinct comet nucleus, is a well known short-period comet of intrinsically low activity with low albedo. The comet was discovered on $5^{th}$ February 1951 by Arend and Rigaux at Uccle, Belgium. The orbital eccentricity is 0.6003 and inclination is 19.04 degrees. The last perihelion of the comet was on $19^{th}$ October 2011 and the next perihelion will be on $15^{th}$ July 2018. Based on the observations, made in the 1980's, and modeling results the estimated mass loss from the comet activity was about 30 kg $s^{-1}$ (Brikett et al., 1987). In the search for surface ice Brooke and Knacke (Brooke and Knacke 1986) reported that no clear evidence was found for surface ice. Following that Veeder et al (Veeder, Hanner et al. 1987) concluded that there is not much ice exposed on the surface, which is rather porous, dark, red and warm. In addition, Millis et al (Millis, A'Hearn et al. 1988) concluded from the combination of results from the nucleus and coma that the nucleus is totally covered with a non-volatile mantle and estimating the active area to be only about 1%. However, the authors suggested that the subsurface ices could be the source for volatile material. In fact, due to the short period, 6.7 years for comet 49P/Arend-Rigaux, repeated processing of the cometary nucleus accompanied with the material loss could have exposed the subsurface layers lately. Therefore, understanding the physical and chemical modifications the cometary nucleus had undergone after every period or a set of periods around the Sun is essential, if we were to develop a complete understanding of a comet. In this letters, we present the results from the recent apparitions since 2006 in combination with laboratory simulation.

## Spitzer spectrum

Spitzer Space Observatory had observed Comet Arend-Rigaux on $7^{th}$ March 2006 using the IRS in staring mode spectroscopy with an AOR ID 16206592. At the time of observations, Comet Arend-Rigaux was at distances of 3.5616 AU from Sun, 3.4289 AU from Spitzer and 3.1332 AU from Earth. We have extracted the spectra following the prescribed procedures. The obtained data sets were reduced using Spectroscopic Modeling Analysis and Reduction Tool (SMART). The spectrum was obtained using the optimal automatic mode of extraction in SMART.

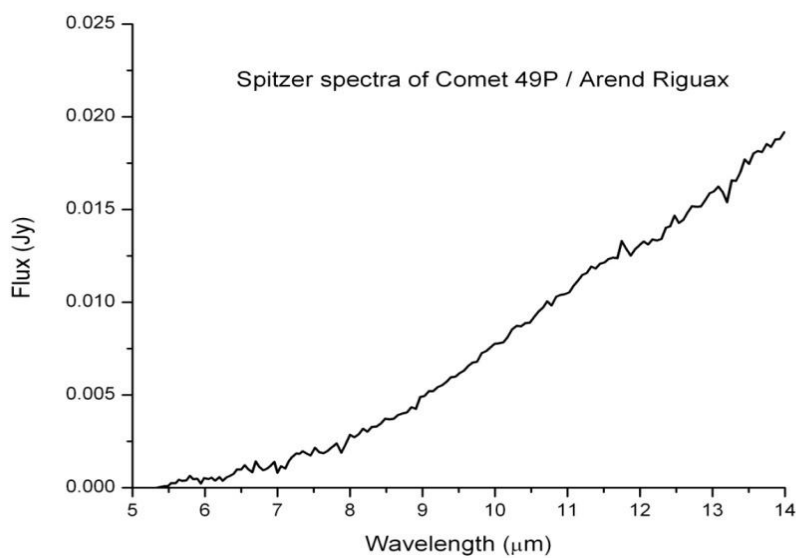

**Figure 1:** Spectra extracted from the 2006 Spitzer observation of comet 49P/Arend-Rigaux. Error values were found to be within ± 10 %.

After data extraction, spectrum shown in Figure 1 was found have weak dip centring around 13 μm. Upon polynomial fit representing the continuum from 8 – 14 μm the dip in the spectra was found to be a prominent broad feature (Figure 2) suggesting an absorption feature in the spectrum due to presence of molecular ices. Though there could be similar absorption features in the 5 – 8 μm spectral region the flux recorded were quite low for any decisive evidences.

**Experimental spectrum**

Laboratory simulation to obtain water ice spectra at 85 K was carried out at the laboratory for low temperature astrochemistry at the Physical Research Laboratory (PRL) (Sivaraman, Mukherjee et al. 2014). The experimental chamber is pumped to <$10^{-9}$ mbar containing a cold head holding zinc selenide (ZnSe) window cooled down to 85 K. Water molecules are introduced into the vacuum chamber condensed on to the cold surface at 85 K. Fourier Transform InfraRed (FTIR) spectrometer operating at mid-infrared region was used to record the InfraRed (IR) spectra. The spectrum recorded was then plotted with the continuum subtracted spectra (Figure 2). In fact, from the spectral comparison, in the 10 – 14 μm, we can conclude that there are exposed water ices on the surface of the comet at the time of observation in 2006. However, the other water absorption around 6 μm could not be compared due to very low flux in the spectra extracted from the Spitzer observation.

From laboratory simulations it is known that amorphous water ice exist at such lower temperatures and a phase transition to crystalline form takes place when the ice is annealed to higher temperatures > 130 K. Conversely, phase transition from crystalline to amorphous water ice at the low temperature is also possible from irradiation processing of the ice (Famá, Loeffler et al. 2010) and references therein). However, Strazzulla et al (Strazzulla, Baratta et al. 1992) discussed the resistant in the irradiation induced phase transition from crystalline to amorphous form for temperatures above 55 K. Based on the above facts and from figure 2, we can even conclude that the water ice exposed as seen during the observation in 2006 could be a recent event due to its amorphous nature. These uncovered ices would be exposed to solar UV photons and thermal processing and therefore could have turned crystalline by the time of the next perihelion approach (October 2011). Further, from previous experimental data (Lowenthal, Khanna et al. 2002) absorption in the 10 – 14 μm spectral region could also correspond to water ice containing isocyanic acid (HNCO).

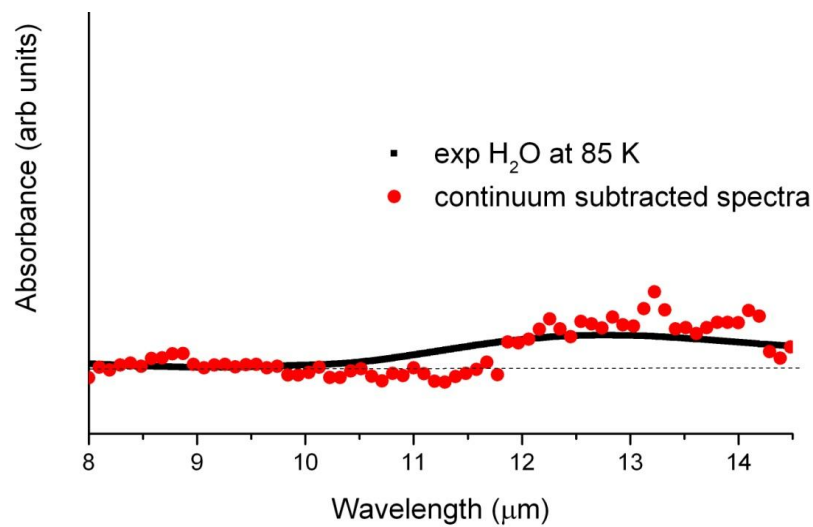

**Figure 2:** IR spectra recorded by depositing water molecules at 85 K compared with the continuum subtracted spectra obtained from Spitzer observation in 2006. Dotted lines indicated zero in the absorbance scale.

**Ground based optical imaging**

We observed the comet on 26$^{th}$ March 2012 from PRL's Gurushikhar Observatory, Mt.Abu India using the 1.2 m telescope. The telescope is equipped with 1296 x 1152 CCD. With on-chip-binning 2x2 pixels, the plate scale is 0.6". The comet was observed using the standard broadband R filter. 30 images of the comet were obtained, each of exposure time 120 seconds. At the time of observations the comet was at a distance of 2.17116AU from Sun and 1.26931 AU from the Earth, at a phase angle of 170.7 deg. All the images were bias subtracted and subsequently flat fielded using twilight sky flats with standard IRAF tools. The flat fielded images were aligned to the photo center of the comet and average combined using IRAF/imcombine.

In order to study the visibility of faint structures of the coma of the comet, an image enhancement technique, the Larson and Sekanina Algorithm (LS) (Larson and Sekanina 1984) of rotational difference was applied to the present set of images. The images were rotated by an angle of 10 degrees in one direction, and then subtracted with the original; then with same rotational shift in the opposite direction and subtracted with the original, and then both the difference images were summed.

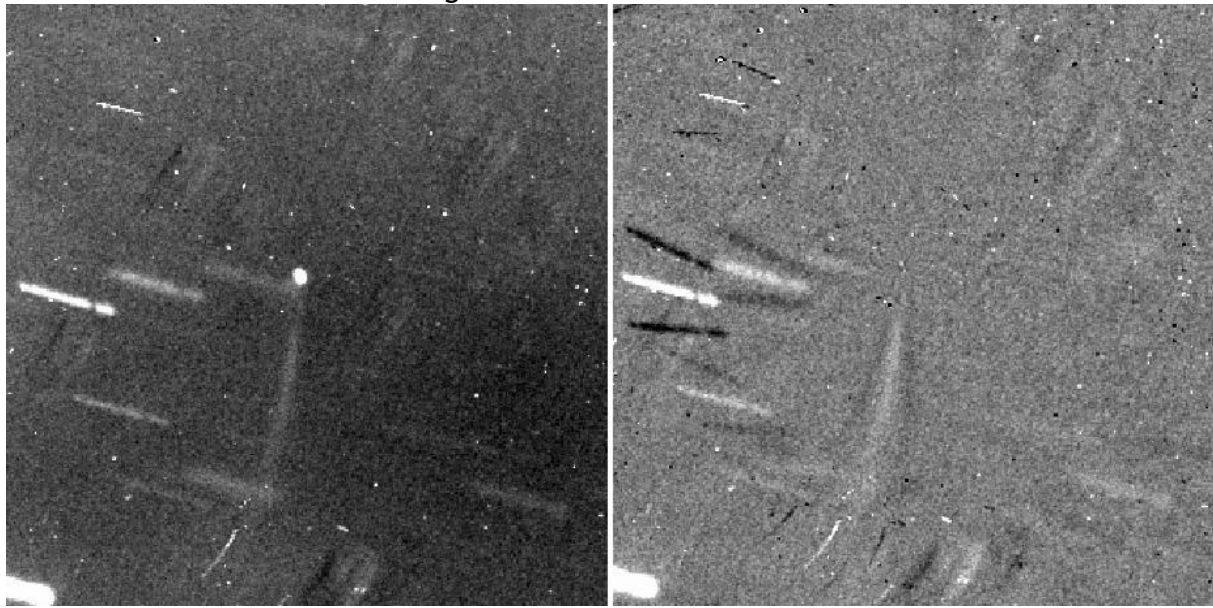

**Figure 3:** (Left) Average combined image of the comet taken in R filter. Exposure Time: 30 x 120 sec . Field of View is 5.6' x 3.7'. (Right) Average combine image, enhanced by LS processing, rotated by an angle of 10 degree. North is to the top and East is to the left in these images.

The comet maintains a stellar profile, but we do see a jet feature in the average combined and LS processed images (Figure 3), which indicates cometary activity. The jet subtends 1.1', which corresponds to 60728 km at the distance of the comet. Note that we do not see any tail in our images. This could be due to low-surface brightness of the tail being swamped by a nearby bright star (BD+31 2316) just off the eastern edge of the field.

The Image processing and photometry was performed in a standard way with the IRAF/APPHOT task. To correct for atmospheric extinction and calibrate the instrumental magnitude of the comet, field stars in the comet frames were used (Table 1). The magnitude of the field stars were determined using the Vizier Sky Tool by overlaying UCAC4 catalog (Zacharias, Finch et al. 2013). Differential photometry was used to calculate the magnitude of the comet.

Table 1: List of the field stars used to calibrate the comet magnitude.

| S.No | RA | DEC | Catalogue magnitude | Instrumental Magnitude |
|---|---|---|---|---|
| 1. | 11h 58m 58.30s | 31d 06m 26.2s | 13.306 | 12.081 |
| 2. | 11h 59m 10s | 31d 06m 09.4s | 14.186 | 12.96 |
| 3. | 11h 59m 17.73s | 31d 09m 26.58s | 13.803 | 12.61 |

To calculate dust production rate the quantity Afρ is used (Equation 1) (Ahearn, Schleicher et al. 1984). This aperture independent quantity is roughly proportional to the dust production rate of a comet assuming equal size distributions of particles in the coma and can be determined from the observations using:

$$Af\rho = (2r\Delta)^2 / \rho * F_{com} / F_{sun} \quad [1]$$

It is defined as the product of albedo (A), the filling factor of the grains in the FOV (f) within an aperture of projected radius ρ, and the projected radius ρ. r [AU] is the comet's heliocentric distance, Δ[cm] is the comet's geocentric distance, ρ [cm] is the aperture radius at the comet distance, $F_{com}$ [erg cm$^{-2}$ s$^{-1}$] is the measured cometary flux in the R filter, and $F_{sun}$ [erg cm$^{-2}$ s$^{-1}$] is the solar flux at 1 AU. Assuming that this model can be used in the present circumstance, we have used an aperture of 10.2" to calculate the corresponding Afρ value. The comet is at 15.88(0.04) mag in the R band and this corresponds to an Afρ value of ~64 cm. This suggests that the comet activity had in fact doubled from more apparitions.

Conclusion

By combining Spitzer observation and laboratory spectra in the 10 μm to 14 μm spectral region we conclude that there were recent exposures of amorphous water ice on the surface of Comet 49P/Arend-Rigaux. In fact, by looking back at the earlier published observations before the Spitzer observation in 2006 we could even propose it was since that orbit in 2006 that the comet had started to be active. This claim holds good due to the comets several orbits around the Sun that had extracted the material of the comet exposing volatile layers beneath. Also the 2011 apparition showing jet activity confirms the comet to have started to be active due to the newly exposed volatile layers, where estimate based on our observations clearly show large amount of material being lost from the comets present active state. In summary, comet 49P/Arend-Rigaux has gone from low active state in the last century to the recent active state since 2006. This reveals the vigorous processing the comet had undergone, as expected for any astronomical object. Our results may find implications in the near future observations to be carried out on Comet 67P/Churyumov-Gerasimenko by Rosetta (and its lander, Philae) and on Comet

C/2013 A1 (Siding Spring) as it emphasizes the need to understand the sub-surface ices in comets. Observations in the next perihelion passage of Comet 49P/Arend-Rigaux in 2018 are encouraged to understand the nature of activity in this comet.

**Acknowledgement**

BS would like to acknowledge the PI Robert Gehrz for the Spitzer observation in 2006. Also BS acknowledges the support from Lydia, Ragavram and K P Subramanian during the course of this work. The authors would like to thank U C Joshi for the discussion.